\documentclass{article}
\usepackage[utf8]{inputenc}
\usepackage[yyyymmdd]{datetime}
\usepackage{url}
\usepackage{tabularx}
\usepackage{tikz}
\usepackage{wasysym}
\usepackage{flowchart}
\usepackage{breakurl}
\usetikzlibrary{shapes,arrows.meta,chains}

\title{A Scalable Architecture for Electronic Payments}
\author{Geoff Goodell, D. R. Toliver, Hazem Danny Nakib}
\date{\textit{\today}}
\def\ccol{0.7em}
\def\smbwd{2cm}
\newcommand{\cz}[1]{\textit{\textbf{#1}}}

\newcolumntype{L}[1]{>{\raggedright\arraybackslash}p{#1}}
\newcolumntype{C}[1]{>{\centering\arraybackslash}p{#1}}
\newcolumntype{R}[1]{>{\raggedleft\arraybackslash}p{#1}}

\newcommand{\ts}{
    \tikzset{>={Latex[width=3mm,length=3mm]}}
    \tikzstyle{line} = [draw, ->, >=latex, ultra thick]
    \tikzstyle{circ} = [
      circle,
      align=center,
      text width=3em,
      text centered,
      inner sep=0mm,
      outer sep=0mm,
      very thick,
      minimum width=0cm,
      minimum height=0cm
    ]
    \tikzstyle{nbox} = [
      draw,
      rectangle,
      thick,
      align=center,
      text width=3.6cm,
      text centered,
      minimum height=1.6em
    ]
    \tikzstyle{n} = [
      rectangle,
      align=left,
      text width=7.2cm,
      minimum width=1.5cm,
      minimum height=1cm
    ]
    \tikzstyle{term} = [
        draw,
        terminal,
        text width=4em,
        text centered,
        minimum width=\smbwd,
        minimum height=0.5cm
    ]
    \tikzstyle{ffbox} = [
        draw,
        rectangle,
        align=center,
        text width=8em,
        text centered,
        minimum width=\smbwd,
        minimum height=1cm
    ]
    \tikzstyle{box} = [
      draw,
      rectangle,
      ultra thick,
      align=center,
      text width=3.2cm,
      text centered,
      minimum height=2.5em
    ]
    \tikzstyle{form} = [
        draw,
        trapezium,
        trapezium left angle=70,
        trapezium right angle=-70,
        align=center,
        text width=5em,
        text centered,
        minimum width=\smbwd,
        minimum height=1cm
    ]
    \tikzstyle{wait} = [
        draw,
        trapezium,
        trapezium left angle=-70,
        trapezium right angle=-70,
        align=center,
        text width=5em,
        text centered,
        minimum width=\smbwd,
        minimum height=1cm
    ]
    \tikzstyle{asset} = [
      box,
      text width=2cm,
      align=center,
      color=white,
      fill=magenta!50!black
    ]
    \tikzstyle{noshape} = [align=center, text centered]
}

\begin{document}

\maketitle




\begin{abstract}

We present a scalable architecture for electronic retail payments via central
bank digital currency and offer a solution to the perceived conflict between
robust regulatory oversight and consumer affordances such as privacy and
control.  Our architecture combines existing work in payment systems and
digital currency with a new approach to digital asset design for managing
unforgeable, stateful, and oblivious assets without relying on either a central
authority or a monolithic consensus system.  Regulated financial institutions
have a role in every transaction, and the consumer affordances are achieved
through the use of non-custodial wallets that unlink the sender from the
recipient in the transaction channel.  This approach is fully compatible with
the existing two-tiered banking system and can complement and extend the roles
of existing money services businesses and asset custodians.

\end{abstract}

\section{Introduction}

We consider the problems posed by modern retail payments in the context of the perceived need for compromise between regulatory compliance and consumer protections.
Retail payments increasingly rely on digital technology,
including both e-commerce transactions via the Internet and in-person
electronic payments leveraging payment networks at the point of sale.  With
cash, customers pass physical objects that are in their possession to merchants.  
In contrast, electronic payments are generally conducted by proxy: Customers
instruct their banks to debit their accounts and remit the funds to
the bank accounts of their counterparties.  For this reason, non-cash retail payments expose customers to a variety of costs and risks, including profiling, discrimination, and value extraction by the custodians of their assets.


A good central bank digital currency (CBDC) would empower individuals 
to make payments using digital objects in their possession rather than accounts that
are linked to their identities, affording them verifiable
privacy and control over their digital payments.  
However, many existing
CBDC proposals require either a centralised system operator or a
global ledger.  Centralised systems entail risks both for the users of
the system as well as for the system operators, 
and global ledgers present
performance bottlenecks as well as an economically inefficient allocation of
transaction costs.


We present a system architecture for retail payments that allows
transactions to take place within a local context, avoiding the problems associated with performance bottlenecks and centralised system operators.
We show how assets that represent obligations
of central banks can be created and exchanged, without requiring a central system operator to process and
adjudicate all of the transactions, 
and without undermining the portability of money throughout the system
or the ability for regulators to ensure compliance.

Although our proposal takes a decentralised approach to processing
transactions, money within our system intrinsically relies upon a trusted
\textit{issuer}.  This could be the central bank itself, but it could also be a
co-regulated federation, such as a national payment network or the operators of
a real time gross settlement system.  Specifically, the issuer is trusted to
oversee the processing of redemptions, wherein CBDC assets are accepted as
valid by their recipients.

Our proposal is fully compatible with the function of existing private-sector
banks. The architecture provides an effective solution for a
variety of different use cases, including those that are sensitive to regulatory compliance requirements, transactional efficiency concerns, or consumer affordances such as privacy and control.
We begin with an examination of the properties required to support such use cases.

The remainder of this article is organised as follows.
Chapter~\ref{s:desiderata} identifies the properties that a payment system
should have as a foundation for a robust set of technical requirements,
Chapter~\ref{s:architecture} specifies the design of our proposed architecture,
Chapter~\ref{s:operational} offers a model for how to deploy and manage a
central bank digital currency (CBDC) system using our architecture,
Chapter~\ref{s:use-cases} describes several use cases that demonstrate the
special capabilities of our proposed design, Chapter~\ref{s:analysis} compares
our design to other payment systems, and Chapter~\ref{s:conclusion} provides a
summary.


\section{Payment system desiderata}
\label{s:desiderata}

To be broadly useful for making payments, and particularly to satisfy the requirements of central bank digital currency, a payment system must have the properties necessary to meet the demands of its use cases. We describe these properties and use cases, and show that they are indeed required.

\newcounter{enumTemp}

\subsection{Asset-level desiderata}
\label{ss:asset}

\begin{itemize}

\item\cz{Integrity.} We say that an asset has \textit{integrity} if it has a
single, verifiable history.  Actors in possession of the asset must be able to
confirm that the asset is genuine and unique; specifically, any two assets that
share any common history must be the same asset.  Desired characteristics of
integrity include:

\begin{enumerate}

\item\textbf{Durability}
\label{des:durability}


Short of stealing the private key of an issuer or breaking the cryptographic
assumptions upon which the system infrastructure depends, it shall not be
possible to create a counterfeit token, it shall not be possible for the party
in possession of a token to spend it more than once, and it shall not be
possible for an issuer to create two identical tokens.  In addition, it shall
not be possible for any actor to mutate the token, once issued.

    
\item\textbf{Self-contained assets}
\label{des:self-contained}


The asset shall be \textit{self-validating}, which is to say that it shall
support a mechanism that allows it to furnish its own proof of integrity, as
part of a process of verifying its authenticity to a recipient or other
interested party.  The purpose of self-validation is to maximise the
flexibility of how assets are used and how risks related to asset ownership and
state can be managed.  In particular, the issuer shall not be required to track
the owner or status of the assets that it has created, and payers shall not
concern themselves with what happens to an asset once it is spent.



\setcounter{enumTemp}{\theenumi}
\end{enumerate}

\item\cz{Control.} An actor has \textit{control} of an asset if that actor and
no other actor possesses the means to specify legitimate changes to the asset,
including features that identify its owner.  Note that \textit{control} implies
the ability to modify the asset in a way that determines the legitimacy of
changes made to the asset by its possessor.  Desired characteristics of control
include:

\begin{enumerate}
\setcounter{enumi}{\theenumTemp}

\item\textbf{Mechanical control}
\label{des:mechanical}

The ability to create a valid transaction is vested in the owner.  No one but the owner can update the state of a specific asset.

\item\textbf{Delegation} 
\label{des:delegation}

The asset owner must be able to retain control of the asset when transferring the responsibility of possession to a custodian. That custodian is then unable to exercise control over that asset, for instance by creating a legitimate update to the asset.  The owner chooses who can exercise control, and owners can delegate possession without delegating control.

\setcounter{enumTemp}{\theenumi}
\end{enumerate}

\item\cz{Possession.} An actor has \textit{possession} of an asset if that
actor and no other actor can effect changes to the asset or reassign possession
of the asset to another actor.  \textit{Possession} implies the ability to deny
possession to others, including the legitimate owner of the asset, on an
incidental or permanent basis (this does not include the possibility for forced legal enforcement to relinquish or return an asset).  In principle, the balance among costs and risks
related to the possession of an asset, including the ability to store assets
safely, can be independently chosen by various actors in the system.  Desired
characteristics of possession include:

\begin{enumerate}
\setcounter{enumi}{\theenumTemp}
\item\textbf{Choice of custodian}
\label{des:custodian}

Asset owners must be able to choose the custodian entrusted with the possession
of their asset.  This contrasts with traditional ledger-based approaches in
which the ledger is the fixed source of truth about an asset and for which an
asset is inextricably bound to that ledger (i.e., moving the asset to another
ledger would involve redemption in the first ledger and a new issuance in the
next ledger).  This property is an essential interoperability feature for any
national currency system.  To mitigate risks such as custodial compromise or
service disruption for sensitive payment systems, asset owners must be able to
choose to have the possession of their assets spread across multiple custodians
(``multiplexing''), such that they require only some portion of them to respond
in order to update the state of their assets. This should be able to occur in a
way that is opaque to the custodians (``oblivious multiplexing''), where each
is concerned only with its own portion of assets which it is providing custody over and is unaware that other custodians are
involved.

\item\textbf{Choice to have no custodian}
\label{des:no-custodian}


The owner of an asset must be able to serve as his or her own custodian.
Specialised custodians are good for mitigating risks, but they always introduce
costs (transaction fees, account fees, latency, and so on) and risks (for
example, intentional or accidental service disruption). To address use cases
that are sensitive to those costs and risks it is necessary to allow
non-specialised actors to also provide custodianship of assets, and in
particular to allow a human owner of an asset to store the asset personally,
using his or her own devices.

\setcounter{enumTemp}{\theenumi}
\end{enumerate}

\item\cz{Independence.} Asset owners shall be free to conduct transactions in the future, with confidence that they will be able to use the assets for the use cases they want.

\begin{enumerate}
\setcounter{enumi}{\theenumTemp}

\item\textbf{Fungibility}
\label{des:fungible}

Each unit is mutually substitutable for each other unit of same issuer, denomination, and vintage, and can be exchanged for cash or central bank reserves.  This is enabled by privacy by design and required for self-determination.

\item\textbf{Efficient lifecycle}
\label{des:efficient-lifecycle}

Transactions must be similar in speed to traditional payment systems, capable
of having near-instant acceptance.  It must be possible for the recipient of an
asset to verify that a transaction is valid and final without the need to
involve a commercial bank at the time of the transaction, and without forcibly
incurring additional costs, risks, or additional technical or institutional
requirements.  Assets must not expire within an unreasonably short timeframe.

\setcounter{enumTemp}{\theenumi}
\end{enumerate}

\end{itemize}

\subsection{System-level desiderata}
\label{ss:system}

\begin{itemize}

\item\cz{Autonomy.} We say that an actor has \textit{autonomy} with respect to
an asset if the actor has both possession and control of the asset and can
modify the asset without creating metadata that can be used to link the actor
to the asset or any specific transaction involving that asset.  The term
\textit{autonomy} is chosen because it reflects the risk that a data subject
might lose the ability to act as an independent moral agent if such records are
maintained~\cite{shaw2017}.  Desired characteristics of autonomy include:

\begin{enumerate}
\setcounter{enumi}{\theenumTemp}

\item\textbf{Privacy by design}
\label{des:privacy-by-design}

The approach must allow users to withdraw money from a regulated entity, such
as a bank or money services business, and then use that money to make payments
without revealing information that can be used to identify the user or the
source of the money.  The assets
themselves, and the transactions in which they are involved, must be untraceable
both to their owners and to other transactions.  The system must be designed to
allow all users to have a sufficiently large anonymity set that they would not
have reason to fear profiling on the part of powerful actors with access to
aggregated data.


\item\textbf{Self-determination for asset owners}
\label{des:self-determination}

Asset owners shall be able to control what they do with assets.  No recipient
can use the system to discriminate against asset owners or impose restrictions
on what a particular owner can do.  Transactions using an asset shall not be
blocked or otherwise flagged by recipients based upon targeting the owner of an
asset or targeting a set of assets associated with some particular transaction
history.

\setcounter{enumTemp}{\theenumi}
\end{enumerate}

\item\cz{Utility.} The system must be generally useful to the public as a means
to conduct most, and perhaps substantially all, retail payments.  Desired
characteristics of utility include:

\begin{enumerate}
\setcounter{enumi}{\theenumTemp}

\item\textbf{Local transactions}
\label{des:local}

It shall be possible to achieve efficient transactions where participants are
able to rely upon local custodians to facilitate acceptance of remittances.
The system shall not rely upon global consensus to determine or verify the
disposition of an asset and shall allow transacting parties to choose an
authority or context that they mutually trust, for example to trust a local
authority in exchange for faster settlement or when access to a wider network
is not possible, without requiring additional trust between counterparties.


\item\textbf{Time-shifted offline transactions}
\label{des:time-shifted}

It shall be possible for a payer to ``time-shift'' third-party trust to achieve
a form of offline payment by first prospectively paying a recipient and then
later, in an offline context, choosing whether to consummate the payment by
selectively revealing additional information.  Time-shifted offline
transactions are akin to purchasing a ticket online and, later, spending it
offline.

\item\textbf{Accessibility}
\label{des:accessibility}

The protocol employed by the system must be accessible and open to all users.
The system must not impose vendor-specific hardware compatibility requirements
and must not require manufacturers of compatible hardware to register with a
central database or seek approval from an authority.  The functionality of the
system must not depend upon trusted computing, secure enclaves, or secure
elements that impose restrictions upon what users can do with their devices.
The system must not require a user to register before acquiring and using a
device, and the possession and use of a physical device must not depend upon a
long-term relationship with a trusted authority, registered business, or asset
custodian.

\setcounter{enumTemp}{\theenumi}
\end{enumerate}

\item\cz{Policy.}  The system must support the establishment of institutional
policies to benefit the public and the national economy.  Desired
characteristics of policy include:

\begin{enumerate}
\setcounter{enumi}{\theenumTemp}

\item\textbf{Monetary sovereignty}
\label{des:monetary}

Monetary sovereignty entails a central bank and government's ability of
controlling the use of the sovereign legal currency within its borders and the
mechanisms within which it is used.  In support of this end, financial
remittances facilitated by the system shall involve direct obligations of the
central bank of the applicable jurisdiction.


\item\textbf{Regulatory compliance}
\label{des:compliance}


The system shall be operated by regulated financial intermediaries that can
establish and enforce rules for their customers.  The system shall provide a
mechanism that would permit financial intermediaries to prove that they have
enforced those rules completely and in every case.  By extension, the system
would allow for the establishment of regulatory requirements for its operators
to support reasonable monitoring by tax authorities for the purpose of
establishing or verifying the income tax obligations of their clients.  Subject
to the limitation that both counterparties to a transaction would not generally
be known, the system would permit system operators to perform analytics on
their customers, for example, by learning the times and size of asset deposits
or withdrawals.  Ideally, the system would also provide a counter-fraud
mechanism by which consumers to verify the validity of merchants.

\end{enumerate}
\end{itemize}

\subsection{Technical requirements}

Next, we translate the asset-level and system-level desiderata into specific
technical and institutional capabilities that are necessary to support a
suitable payment system.  We begin by identifying the technical requirements
for an institutionally supportable digital currency that supports verifiable
privacy for consumers, wherein consumers are not forced to rely upon promises
by trusted actors:

\begin{itemize}

\item\cz{Blind signatures.}  Consumer agents must implement \textit{blinding}
and \textit{unblinding} with semantics similar to the blind signatures proposed
by Chaum in his original article~\cite{chaum1982} and further elaborated in his
more recent work with the Swiss National Bank~\cite{chaum2021}.  Specifically,
it must be possible for users to furnish a block of data to an issuer, ask the
issuer to sign it, then transform the response into a valid signature on a new
block of data that the issuer has never seen before and cannot link to the
original block of data.  This allows transactions that do not link the identity
of the sender to the identity of the recipient, as a way to achieve privacy by
design (\ref{des:privacy-by-design}).

\item\cz{Distributed ledger.} Participants in a clearing network overseen by a
central bank must have access to a suitable distributed ledger technology (DLT)
system~\cite{iso22739} that enables them to collectively maintain an immutable
record that can be updated with sufficient frequency to provide transaction
finality that is at least as fast as domestic bank wires. This helps ensure
both durability of assets (\ref{des:durability}) and self-determination for
users (\ref{des:self-determination}) as described in
Section~\ref{s:desiderata}.

\item\cz{Open architecture.} The system must fully support the semantics for
digital currency specified by Goodell, Nakib, and Tasca~\cite{goodell2021}.
Specifically, we assume that retail users of digital currency have access to
non-custodial wallets that satisfy certain privacy and accessibility
requirements described in Section~\ref{ss:system}, specifically requirements
(\ref{des:no-custodian}), (\ref{des:privacy-by-design}),
(\ref{des:self-determination}), and (\ref{des:accessibility}).

\item\cz{Fungible tokens.} The digital currency tokens themselves must satisfy
the fungibility requirement (\ref{des:fungible}) described in Section~\ref{ss:asset}.

\item\cz{Institutional controls.}  System operators must possess capabilities
that support the policy requirements described in Section~\ref{ss:system},
specifically requirements (\ref{des:monetary}) and (\ref{des:compliance}).

\end{itemize}

Moving to a digital form of currency brings a variety of potential benefits when compared to paper currency, including cryptographic signatures, cryptographic shielding, flexible semantics, reduced management costs, and being able to efficiently transfer units of currency over large distances. 

However, it is also important to re-capture some of the benefits of physical currency.
In order to have \textit{self-contained assets} with \textit{custodial choice}, we
need a representation for our assets that is unforgeable, stateful, and
oblivious:

\begin{itemize}

\item\cz{Unforgeable.} Every asset must be unique, and it can only be created
once.  No set of adversarial actors can repeat the process of creating an asset
that has already been created.  Note that this requirement is different than a
"globally unique identifier", which is merely unlikely to be reused by an
honest actor, but which any adversarial actor can reuse for any other asset.
True unforgeability requires that once an asset is created, it is impossible to reuse its identifier for any another asset.  This property is required for
durability (\ref{des:durability}),
custodial choice (\ref{des:custodian}),
the choice to have no custodian (\ref{des:no-custodian}),
local transactions (\ref{des:local}), and
time-shifted offline transactions (\ref{des:time-shifted}).

\item\cz{Stateful.} Every asset has its own independent state, and as the state
of an asset changes over time, the asset remains unique and unforgeable.  No
set of adversarial actors, including non-issuer owners, can create a second
version of the asset with a different state.  Note that this requirement
precludes using any kind of ``access control token'', such as an HMAC, signed
attestation, or even a blinded signature scheme asset, which cannot accumulate state over time and must be returned in precisely the same form as created.  
The requirements of self-contained assets (\ref{des:self-contained}),
mechanical control (\ref{des:mechanical}),
and delegation (\ref{des:delegation}) necessitate that assets maintain their own state.

\item\cz{Oblivious.} Once finality is achieved following the transfer of an
asset to a new owner all of the previous owners, including the issuer, have no
obligation to know any aspect of its
future state changes and transfers.  There is no residual
risk to the new owner that the transaction will be undone by either a previous owner or the system itself.
Note that encryption does not suffice: there must be no requirement to inform
previous owners that state changes have occurred, and previous owners must not
be required to do any extra work to accommodate those changes.  Otherwise, the
self-determination (\ref{des:self-determination}) and efficient lifecycle (\ref{des:efficient-lifecycle}) requirements would be compromised.  


Paper bank notes are a good example of obliviousness.
No entity knows where every bank note is, or what everyone's billfolds hold. If
anyone, including the mint, were guaranteed to know this information, then it
would prevent paper money from being useful in many of its required use cases.
Although obliviousness and privacy are closely related, obliviousness is really
about efficiency: It is acceptable for the mint to know where some bills are
and the contents of some billfolds.

\end{itemize}


These qualities combine together to provide assets, referred to as \textit{USO
assets} in this document, that have very similar qualities to paper currency.
While assets embodying these qualities are not readily available at this time,
this is an area of active study and promising results. Given such assets in
combination with the technologies mentioned above, our architecture is able to
fulfill the complete list of requirements for a payment system. In particular,
CBDC created using our architecture can meet the use case demands of paper
currency as well as the demands of electronic payment systems in a single
architecture, without requiring trusted hardware or heavyweight consensus
systems.

The requirement for a USO asset to be stateful means it must be able to prove
its state has finality. The requirement for a USO asset to be oblivious means
that the asset must carry a \textit{proof of provenance} (POP) that allows it
to demonstrate its validity on its own, as no other part of the system is
required to have it. The requirement for a USO asset to be unforgeable means
this proof carries the same weight as if it came from directly the issuer
itself, so the issuer acts as the \textit{integrity provider} of the POP.

Obliviousness implies there can be other systems between the asset owner and
the integrity provider. These systems serve as \textit{relays} in the creation
of the POP. Relays are common carriers, like network carriers. In fact a relay
knows considerably less than a network carrier: it accepts hashes, and emits
hashes of those hashes, and by design is completely oblivious to everything
else.

\section{An efficient, general-purpose architecture for CBDC}
\label{s:architecture}

\noindent 


In this section we propose a method for creating a retail central bank digital
currency (CBDC) that supports private payments wherein the owner maintains
custody of her digital assets.  It achieves the necessary properties for a
general purpose payment system described in the previous section by extending
the approach proposed by Goodell, Nakib, and Tasca~\cite{goodell2021} with a
new asset model that eliminates the need for global consensus with regard to
every transaction.  While our new approach requires that the central bank must
operate some real-time infrastructure, we show that this requirement can be
addressed with a lightweight, scalable mechanism that mitigates the risk to
resilience and operational security.

Suppose that a user, Alice, wants to withdraw retail CBDC for her
general-purpose use in making retail payments.  We assume that the recipient of
any payment that Alice makes will require one or more valid tokens from a
trusted issuer $I$ containing content $k$ that has been signed using signature
function $s(k, I)$. We further assume, following the arguments made in earlier
proposals for privacy-preserving retail CBDC~\cite{goodell2021,chaum2021}, that
she will be able to use a \textit{blinding} function $b$, known only to Alice,
to request a blind signature on $b(k)$ to which she can apply an
\textit{unblinding} function $b^{-1}$, also known only to Alice, to reveal the
required signature:

\begin{equation}
b^{-1}(s(b(k), I))=s(k, I)
\end{equation}


The signature $s(k, I)$ appearing at the beginning of a USO asset's history
shows that it was generated correctly by the CBDC's issuer or by one of its
delegates, which we shall call \textit{minters}.  Minters are subject to a
\textit{minting invariant} wherein every time a minter satisfies a request for
a set of signatures of a particular value, it must also cancel a corresponding
set of CBDC assets of equal value, and vice-versa.  The function of a minter,
therefore, is to \textit{recycle} CBDC, and not to issue or destroy it.

The proof of provenance of a USO asset allows its recipient to verify that it
has the same integrity as if it were in the issuer's database.  These proofs of
provenance are a powerful enabling feature for a retail CBDC, since assets can
be transacted without the need to maintain accounts.  Additionally, the
expected costs of operating the issuer's infrastructure is much smaller at
scale than the costs associated with operating traditional distributed ledger
infrastructures in which the record of each transaction is maintained in a
global ledger.

However, unlike transferring blinded assets in a classical ledger system,
whether distributed~\cite{goodell2021} or not~\cite{chaum2021}, transferring
USO assets from one party to another explicitly leaves behind an audit trail
that can be used by the bearers of an asset to recognise the asset when it is
inspected, transacted or seen in the future.  A USO asset's proof of provenance
is permanently updated each time it is transferred to a new recipient.  If the
same asset were to be associated with multiple transactions, then a single
party to any of the transactions would be able to recognise the asset across
all of its transactions, which could potentially compromise the privacy of the
other parties.

\begin{figure}[ht]
\begin{center}
\hspace{-0.8em}\scalebox{1.2}{
\begin{tikzpicture}[>=latex, node distance=3cm, font={\sf \small}, auto]\ts
\tikzset{>={Latex[width=4mm,length=4mm]}}
\node (w1) at (0, 0) [] {
    \scalebox{0.08}{\includegraphics{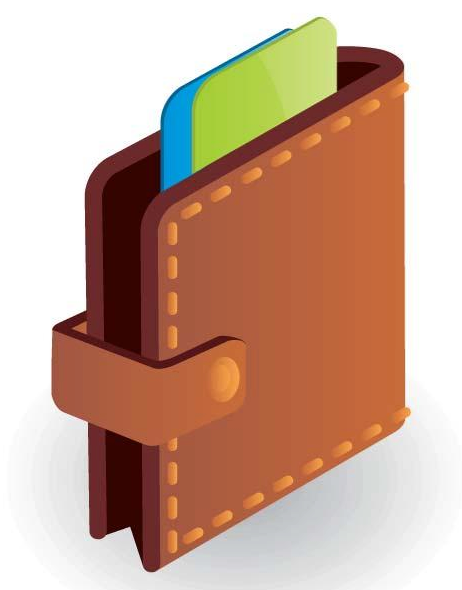}}
};
\node (b1) at (4, 3) [] {
    \scalebox{2}{\includegraphics{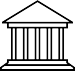}}
};
\node (b2) at (8, 3) [] {
    \scalebox{0.6}{\includegraphics{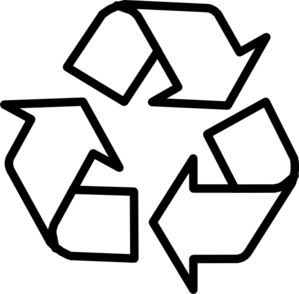}}
};
\node (s1) at (4, -3) [] {
    \scalebox{2}{\includegraphics{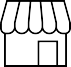}}
};
\node (r1) at (8, -3) [] {
    \scalebox{0.3}{\includegraphics{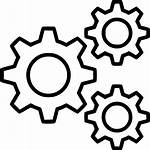}}
};
\node (t1) at (0.4, 2.7) [box, text width=1.5cm, align=center, fill=green!30] {
    signature request \& payment
};
\node (t2) at (6, 4.4) [box, text width=1.5cm, align=center, fill=green!30] {
    signature request \& payment
};
\node (t3) at (6, 1.8) [box, text width=1.5cm, align=center, fill=orange!30] {
    blind signature
};
\node (t4) at (3.5, 0.5) [box, text width=1.5cm, align=center, fill=orange!30] {
    blind signature
};
\node (t5) at (0.1, -2.5) [asset] {asset data};
\node (t6) at (6, -1.8) [box, text width=2cm, align=center, fill=magenta!30] {
    xfer asset to merchant
};
\node (t7) at (6, -4.2) [box, text width=2cm, align=center, fill=blue!30] {
    proof of provenance
};
\node (c1) at (-1.5, 0) [noshape, text width=2cm, align=center] {consumer wallet};
\node (c2) at (4, 4.2) [noshape, text width=1.5cm, align=center] {commercial bank};
\node (c3) at (9.3, 3) [noshape, text width=1.5cm, align=center] {minter};
\node (c4) at (4, -3.8) [noshape, text width=1.5cm, align=center] {merchant};
\node (c5) at (9.2, -3) [noshape, text width=1.5cm, align=center] {relay};
\draw[->, line width=1mm] (w1) edge[bend left=20] (b1);
\draw[->, line width=1mm] (b1) edge[bend left=20] (b2);
\draw[->, line width=1mm] (b2) edge[bend left=20] (b1);
\draw[->, line width=1mm] (b1) edge[bend left=20] (w1);
\draw[->, line width=1mm] (w1) edge[bend right=20] (s1);
\draw[->, line width=1mm] (s1) edge[bend left=20, color=red] (r1);
\draw[->, line width=1mm] (r1) edge[bend left=20, color=red] (s1);
\end{tikzpicture}}
\end{center}

\caption{\cz{Schematic representation of the CBDC journey from the perspective
of a consumer.}}

\label{f:consumer-lifecycle}
\end{figure}
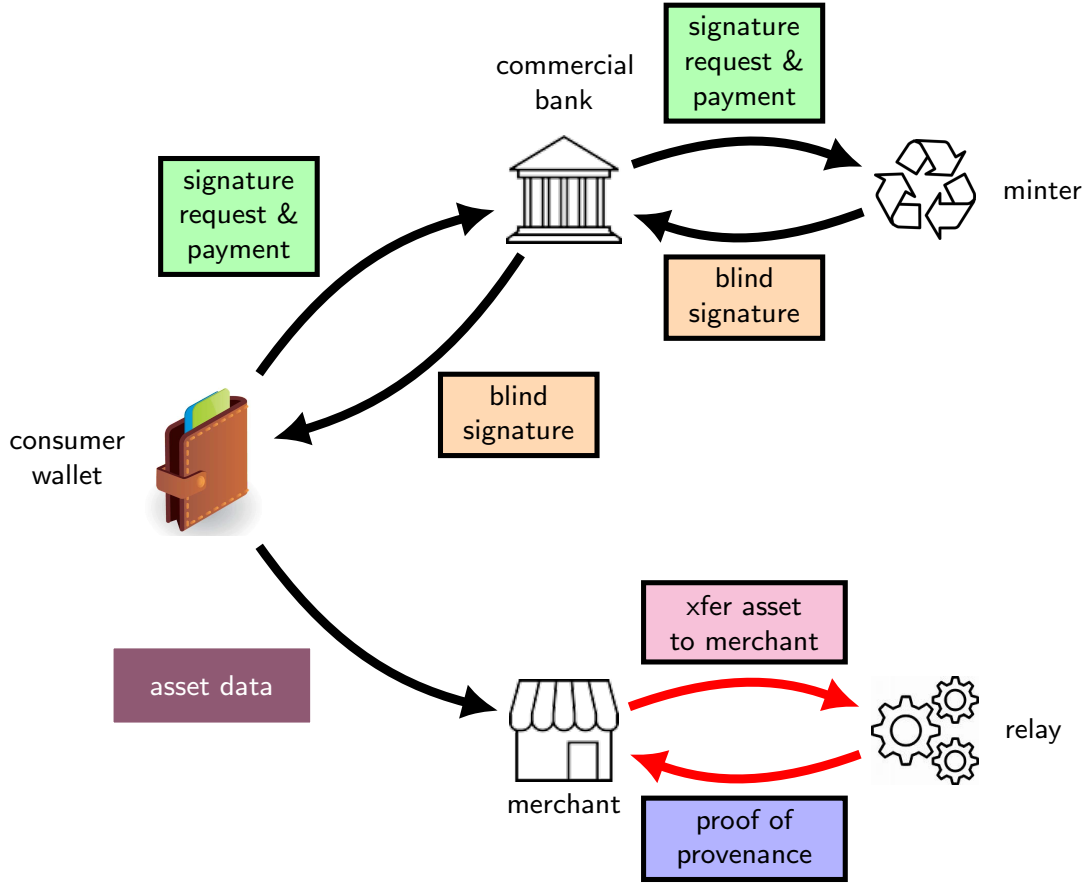

It follows that if Alice wants an asset that she can spend privately, she must
create it herself. Alice establishes her own USO asset privately, and
subsequently populates it with the signature $s(k, I)$.  Having done this she
can then safely transfer the asset to Bob without concern.
Figure~\ref{f:consumer-lifecycle} provides a visualisation of the CBDC journey
from the perspective of a consumer.

Once Bob receives the asset from Alice, he has a choice.  One option is to
transfer it to a bank, perhaps to deposit the proceeds into his account with
the bank, or to request a freshly minted CBDC asset as Alice had done earlier.
If he chooses to deposit the proceeds into his account, then the bank now has a
spent CBDC asset that it can exchange for central bank reserves or use to
satisfy requests for new signed CBDC assets from its other account holders.
Alternatively, Bob could transfer the CBDC onward without returning it to the
bank, bearing in mind that Bob would not be anonymous when he does; see
Section~\ref{ss:chained} for details.

We organise Alice's engagement lifecycle with the asset in a five-step process,
as shown in Figure~\ref{f:aliceflow}:

\begin{figure}[ht]
\begin{center}
\sf\begin{tikzpicture}[>=latex, node distance=3cm, font={\sf \small}, auto]\ts
\tikzset{>={Latex[width=2mm,length=2mm]}}
\node (terminal1) at (0,0.0) [term] {START};
\node (form1) at (0,-1.3) [form, text width=9.3em] {
    (1) Create asset $F_0=\{A, G_0, s((d,I_d), I)\}$
};
\node (box1) at (0,-2.9) [ffbox, text width=10em] {
    (2) Request signature $s(b(F_0), I_d)$ from issuer $I$
};
\node (wait2) at (0,-4.5) [wait, text width=3.4em] {Wait $dt$};
\node (form2) at (0,-6.3) [form, text width=9.4em] {
    (3) Create update $F_1$ to add signature $s(F_0, I_d)$ and transfer asset to $B$
};
\node (box2) at (0,-8.1) [ffbox, text width=10em] {
    (4) Send $F_0$ and $F_1$ to relay $G$
};
\node (box3) at (0,-9.6) [ffbox, text width=10em] {
    (5) Furnish proof to owner of $B$
};
\node (terminal2) at (0,-10.9) [term] {END};
\draw[->] (terminal1) -- (form1);
\draw[->] (form1) -- (box1);
\draw[->] (box1) -- (wait2);
\draw[->] (wait2) -- (form2);
\draw[->] (form2) -- (box2);
\draw[->] (box2) -- (box3);
\draw[->] (box3) -- (terminal2);
\end{tikzpicture}
\rm\vspace{-1em}
\end{center}

\caption{\cz{Typical consumer engagement lifecycle.} Parallelograms represent
USO asset operations.}

\label{f:aliceflow}
\end{figure}
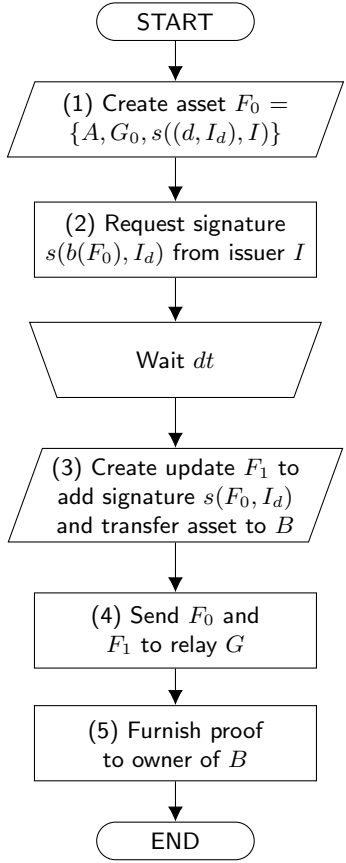

\begin{enumerate}

\item First, Alice chooses a service provider that maintains a relay $G$, and
creates a new USO asset that refers to some specific prior commitment $G_0$
published by the relay.  For each CBDC token that Alice wishes to obtain, she
generates a new pair of keys using asymmetric cryptography and embeds the
public key $A$ and $G_0$ along with the public key of the proposed digital
currency issuer $I$, the denomination $d$, and a certificate $s((d,I_d), I)$
containing the key used by the issuer to sign tokens of denomination $d$ into a
template for a new, unique update $F_0=\{A, G_0, s((d, I_d), I)\}$ as the
foundation for a new asset $F$.  Note that for Alice to ensure that her
subsequent spending transactions are not linked to each other, she must repeat
this step, creating a new key pair for each asset that she wants to create, and
optionally choosing different values for the other parameters as well.

\item Next, Alice creates $b(F_0)$ using blinding function $b$ and sends it to
her bank along with a request for a blind signature from a minter using the key
for the correct denomination $I_d$, which in the base case we assume to be the
central bank.  Alice is effectively requesting permission to validate asset $F$
as legitimate national digital currency (the sovereign legal tender within that
jurisdiction), so, presumably, the bank will require Alice to provide
corresponding funds, such as by providing physical cash, granting the bank
permission to debit her account, or transferring digital currency that she had
previously received in the past.  See Figure~\ref{f:step-2}.  Alice's bank
shall forward her request $b(F_0)$ to the central bank along with central bank
money (cash, central bank reserves, or existing CBDC assets) whose total value
is equal to the value of the CBDC that Alice is requesting.  The bank shall
then provide Alice with the signature $s(b(F_0), I_d)$.

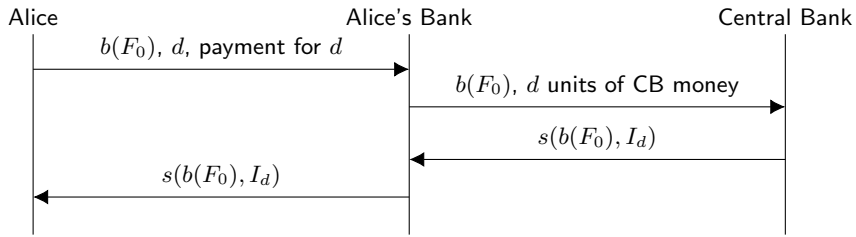
\begin{figure}[ht]
\begin{center}
\sf\begin{tikzpicture}[>=latex, node distance=3cm, font={\sf \small}, auto]\ts
\tikzset{>={Latex[width=2mm,length=2mm]}}
\node (n1) at (0,0) {Alice};
\node (n2) at (5,0) {Alice's Bank};
\node (n3) at (10,0) {Central Bank};
\draw (n1) -- (0,-2.9) {};
\draw (n2) -- (5,-2.9) {};
\draw (n3) -- (10,-2.9) {};
\draw[->] (0,-0.7) -- node[above] {
    $b(F_0)$, $d$, payment for $d$
} (5,-0.7);
\draw[->] (5,-1.2) -- node[above] {
    $b(F_0)$, $d$ units of CB money
} (10,-1.2);
\draw[->] (10,-1.9) -- node[above] {
    $s(b(F_0), I_d)$
} (5,-1.9);
\draw[->] (5,-2.4) -- node[above] {
    $s(b(F_0), I_d)$
} (0,-2.4);
\end{tikzpicture}
\rm\vspace{-1em}
\end{center}

\caption{\cz{Protocol for Step 2.} The validation of $d$ units of digital
currency.}

\label{f:step-2}
\end{figure}

\item At this point, Alice can now ``unblind'' the signature received from the
minter to yield $s(F_0, I_d)$, which is all that is required to create valid
CBDC.  To mitigate the risk of timing attacks that could be used to correlate
her request for digital currency with her subsequent activities, Alice should
wait for some period of time $dt$, before conducting a transaction with the
valid CBDC received as well as before sharing the unblinded signature $s(F_0,
I_d)$.  Alice's privacy derives from the number of tokens that are
``in-flight'' (outstanding) at any given moment.  If she transacts too quickly
after completing her withdrawal, then her spending transaction might be traced
to her withdrawal.

When Alice is ready to conduct a transaction with Bob, she creates a new update
$F_1$ wherein she updates the metadata of $F$ to include the signature $s(F_0,
I_d)$ and transfer ownership to Bob using his public key $B$.  Optionally,
Alice might want to confirm that $B$ legitimately belongs to Bob's business, in
which case Bob could furnish a certificate for his public key.  We also imagine
that regulators might impose additional requirements that would apply at this
stage, which we describe in Section~\ref{ss:regulatory}.  Observe that neither
the asset $F_0$ nor its update $F_1$ contain any information about Alice, her
wallet, or any other assets or transactions.

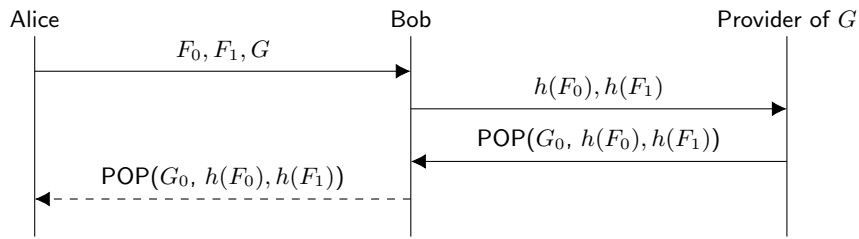
\begin{figure}[ht]
\begin{center}
\sf\begin{tikzpicture}[>=latex, node distance=3cm, font={\sf \small}, auto]\ts
\tikzset{>={Latex[width=2mm,length=2mm]}}
\node (n1) at (0,0) {Alice};
\node (n2) at (5,0) {Bob};
\node (n3) at (10,0) {Provider of $G$};
\draw (n1) -- (0,-2.9) {};
\draw (n2) -- (5,-2.9) {};
\draw (n3) -- (10,-2.9) {};
\draw[->] (0,-0.7) -- node[above] {
    $F_0,F_1,G$
} (5,-0.7);
\draw[->] (5,-1.2) -- node[above] {
    $h(F_0), h(F_1)$
} (10,-1.2);
\draw[->] (10,-1.9) -- node[above] {
    POP($G_0$, $h(F_0), h(F_1)$)
} (5,-1.9);
\draw[->, dashed] (5,-2.4) -- node[above] {
    POP($G_0$, $h(F_0), h(F_1)$)
} (0,-2.4);
\end{tikzpicture}
\rm\vspace{-1em}
\end{center}

\caption{\cz{Protocol for Step 4, Option 1.} Alice gives Bob possession and
control, and Bob registers the update.}

\label{f:step-4-1}
\end{figure}

\begin{figure}[ht]
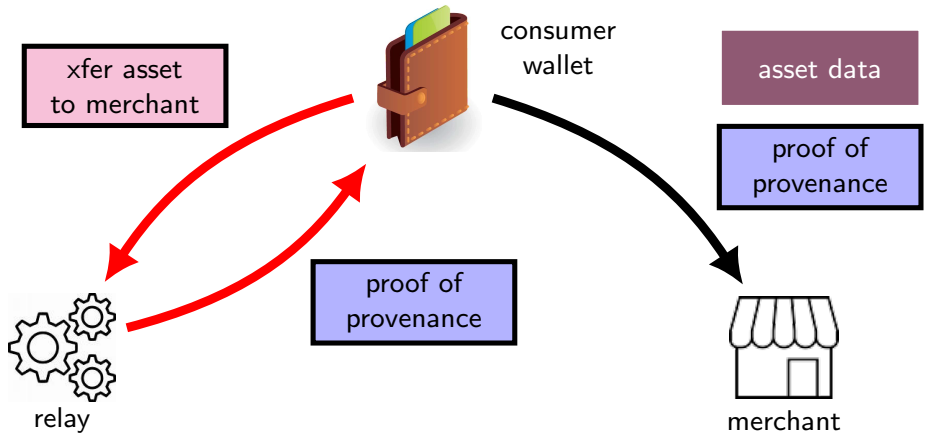

\begin{center}
\hspace{-0.8em}\scalebox{1.2}{
\begin{tikzpicture}[>=latex, node distance=3cm, font={\sf \small}, auto]\ts
\tikzset{>={Latex[width=4mm,length=4mm]}}
\node (w1) at (0, 0) [] {
    \scalebox{0.08}{\includegraphics{images/wallet-vector-xp.png}}
};
\node (s1) at (4, -3) [] {
    \scalebox{2}{\includegraphics{images/s-shop.png}}
};
\node (r1) at (-4, -3) [] {
    \scalebox{0.3}{\includegraphics{images/gears.jpg}}
};
\node (t5) at (-3.3, -0.1) [box, text width=2cm, align=center, fill=magenta!30] {
    xfer asset to merchant
};
\node (t6) at (4.4, -1.0) [box, text width=2cm, align=center, fill=blue!30] {
    proof of provenance
};
\node (t7) at (4.4, 0.1) [asset] {asset data};
\node (t7) at (-0.1, -2.5) [box, text width=2cm, align=center, fill=blue!30] {
    proof of provenance
};
\node (c1) at (1.5, 0.3) [noshape, text width=2cm, align=center] {consumer wallet};
\node (c4) at (4, -3.8) [noshape, text width=1.5cm, align=center] {merchant};
\node (c5) at (-4, -3.8) [noshape, text width=1.5cm, align=center] {relay};
\draw[->, line width=1mm] (w1) edge[bend right=20, color=red] (r1);
\draw[->, line width=1mm] (r1) edge[bend right=20, color=red] (w1);
\draw[->, line width=1mm] (w1) edge[bend left=20] (s1);
\end{tikzpicture}}
\end{center}

\caption{\cz{Schematic representation of Step 4, Option 2.}}

\label{f:alt-journey}
\end{figure}

\begin{figure}[ht]
\begin{center}
\sf\begin{tikzpicture}[>=latex, node distance=3cm, font={\sf \small}, auto]\ts
\tikzset{>={Latex[width=2mm,length=2mm]}}
\node (n1) at (0,0) {Alice};
\node (n2) at (5,0) {Provider of $G$};
\node (n3) at (-5,0) {Bob};
\draw (n1) -- (0,-3.1) {};
\draw (n2) -- (5,-3.1) {};
\draw (n3) -- (-5,-3.1) {};
\draw[->] (0,-0.7) -- node[above] {
    $h(F_0), h(F_1)$
} (5,-0.7);
\draw[->] (5,-1.4) -- node[above] {
    POP($G_0$, $h(F_0), h(F_1)$)
} (0,-1.4);
\draw[->] (0,-1.9) -- node[above] {
    $F_0, F_1, G$
} (-5,-1.9);
\draw[->] (0,-2.6) -- node[above] {
    POP($G_0$, $h(F_0), h(F_1)$)
} (-5,-2.6);
\end{tikzpicture}
\rm\vspace{-1em}
\end{center}

\caption{\cz{Protocol for Step 4, Option 2.} Alice registers the update
herself, giving Bob control first and possession later.}

\label{f:step-4-2}
\end{figure}
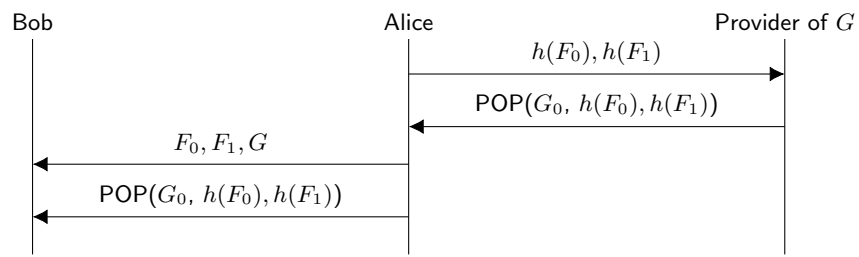

\item To consummate the transaction, $h(F_0)$ and $h(F_1)$ must be sent to
relay $G$, wherein $h$ is a selector function that can be used to demonstrate
that Alice had committed to creating the asset $F_0$ and its update $F_1$,
respectively.  In particular, $h$ may be a hash function.  Alice has two
options for how to proceed:

\begin{itemize}

\item\cz{(Option 1.)} Alice sends the identity of the
relay $G$ along with the asset $F_0$ and its update $F_1$ to Bob (see
Figure~\ref{f:step-4-1}), and Bob sends $h(F_0)$ and $h(F_1)$ to the relay.  At
this point, Bob may furnish the POP of the transaction to Alice, once he
receives it, as a receipt.

\item\cz{(Option 2.)} Alice sends $h(F_0)$
and $h(F_1)$ to the relay directly and subsequently furnishes the asset and its
proof of provenance to Bob (see Figures~\ref{f:alt-journey}
and~\ref{f:step-4-2}).

\end{itemize}

\item Finally, if Alice had chosen Option 2 for the previous step, then she
should reveal to Bob the POP indicating that the transaction is done.  If Alice
had chosen Option 1 for the previous step, then Bob will be able to verify this
himself.

\end{enumerate}

Note that once Alice has transferred the CBDC asset to Bob, nothing about the
asset or its proofs of provenance can be used to link the asset to Alice, her
devices, or her other transactions, regardless of what Bob does with the asset
going forward.  Broadly speaking, these are the same protections that Alice has
when she uses cash, although we expect that regulated financial intermediaries
will generally always learn that Bob receives a CBDC asset when Bob receives an
asset from a non-custodial wallet.

Our architecture provides a general framework for specifying which assets are
considered valid.  Importantly, and unlike some digital currency system
designs, our system allows all of the rules to be implemented at the edge
rather than inside the network itself.  For example, because a regulated
financial intermediary has a role in every transaction, a bank accepting CBDC
assets as deposits might implement a rule requiring that an asset must have
been previously transacted at most once.

Alice's privacy depends upon Alice not binding her identity to the transaction
in some way, for example by embedding her personal information into a
transaction or by linking the transaction to a wallet identifier.  In all
cases, we expect that only the initial consumer, Alice, enjoys the benefits of
consumer protection.  Subsequent recipients of an asset do not have such
protections, and rules enforced by banks that receive assets can impose
explicit requirements on all of the participants in a chain of transactions.
Note that a point of trust is required for any fair transaction between two
untrusting parties~\cite{pagnia1999}.

\section{Operational considerations}
\label{s:operational}

Although our architecture could be applied to arbitrary digital currency
applications, including digital currency and e-money issued by private-sector
banks, we assume that this architecture is most useful for the implementation
of central bank digital currency (CBDC), wherein central banks would be the
issuers of currency for use by the general public to facilitate payments in
domestic retail contexts.  CBDC would represent part of the monetary base (M0),
like cash and central bank reserves.

In this section, we consider operational concerns for the various parties
involved in a CBDC distribution, including central banks, private-sector banks,
clearinghouses, merchants, and consumers.  In particular, we show that the
system is able to support lightweight requirements for central banks as well as
for end-user devices, including both mobile wallets for consumers and merchant
devices at the point-of-sale.

\subsection{Operational model}

\begin{figure}[ht]
\begin{center}
\hspace{-0.8em}\scalebox{0.9}{
\begin{tikzpicture}[>=latex, node distance=3cm, font={\sf \small}, auto]\ts
\tikzset{>={Latex[width=4mm,length=4mm]}}
\node (w1) at (0, 0) [] {
    \scalebox{0.08}{\includegraphics{images/wallet-vector-xp.png}}
};
\node (b1) at (4, 6) [] {
    \scalebox{2}{\includegraphics{images/s-sender-bank.png}}
};
\node (b2) at (12, 6) [] {
    \scalebox{2}{\includegraphics{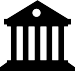}}
};
\node (s1) at (4, -6) [] {
    \scalebox{2}{\includegraphics{images/s-shop.png}}
};
\node (r1) at (3.6, -1.8) [] {
    \scalebox{0.3}{\includegraphics{images/gears.jpg}}
};
\node (r2) at (3.6, 1.8) [] {
    \scalebox{0.3}{\includegraphics{images/gears.jpg}}
};
\node (r3) at (7.2, -1.8) [] {
    \scalebox{0.3}{\includegraphics{images/gears.jpg}}
};
\node (r4) at (7.2, 1.8) [] {
    \scalebox{0.3}{\includegraphics{images/gears.jpg}}
};
\node (c0) at (5.5,0) [
    circ,
    draw,
    line width=1.2mm,
    color=orange,
    minimum height=3.5cm,
    minimum width=3.5cm
] {};
\node (c1) at (5.5,0) [circ, draw, thick, minimum height=3.4cm, minimum width=3.4cm] {};
\node (c2) at (5.5,0) [circ, draw, thick, minimum height=3.6cm, minimum width=3.6cm] {};
\node (p1) at (5.5,0) [align=center] {\textbf{Integrity}\\DLT System\\(relays)};
\node (c3) at (12.5,0) [
    circ,
    draw,
    line width=1.2mm,
    color=orange,
    minimum height=3.5cm,
    minimum width=3.5cm
] {};
\node (c4) at (12.5,0) [circ, draw, thick, minimum height=3.4cm, minimum width=3.4cm] {};
\node (c5) at (12.5,0) [circ, draw, thick, minimum height=3.6cm, minimum width=3.6cm] {};
\node (p2) at (12.5,0) [align=center] {\textbf{Monitoring}\\DLT System\\(minters)};
\node (x1) at (10.6, -1.8) [] {
    \scalebox{0.6}{\includegraphics{images/recycle.png}}
};
\node (x2) at (10.6, 1.8) [] {
    \scalebox{0.6}{\includegraphics{images/recycle.png}}
};
\node (x3) at (14.2, -1.8) [] {
    \scalebox{0.6}{\includegraphics{images/recycle.png}}
};
\node (x4) at (14.2, 1.8) [] {
    \scalebox{0.6}{\includegraphics{images/recycle.png}}
};
\node (b3) at (12, -6) [] {
    \scalebox{2}{\includegraphics{images/s-sender-bank.png}}
};
\node (i1) at (-1.5, 0) [noshape, text width=2cm, align=center] {consumer wallet};
\node (i2) at (2.4, 6) [noshape, text width=1.8cm, align=center] {commercial bank};
\node (i3) at (13.3, 6) [noshape, text width=1.5cm, align=center] {central bank};
\node (i4) at (2.6, -6) [noshape, text width=1.8cm, align=center] {merchant};
\node (i5) at (13.6, -6) [noshape, text width=1.8cm, align=center] {commercial bank};
\draw[->, line width=1mm] (b1) edge[bend right=20] node[above, sloped] {withdraw} (w1);
\draw[<->, line width=1mm] (b1) edge[bend left=20] node[above, sloped] {
    issue, destroy
} (b2);
\draw[->, line width=1mm] (x2) edge[bend right=20] node[above, sloped] {recycle} (b1);
\draw[->, line width=1mm] (b3) edge[bend right=20] node[above, sloped] {recycle} (x1);
\draw[<->, line width=1mm] (r3) edge[bend left=20, color=red] (x1);
\draw[->, line width=1mm] (w1) edge[bend right=20] node[above, sloped] {spend} (s1);
\draw[<->, line width=1mm] (s1) edge[bend right=20, color=red] (r1);
\draw[<->, line width=1mm] (b1) edge[bend left=20, color=red] (r2);
\draw[<->, line width=1mm] (b3) edge[bend left=20, color=red] (r3);
\draw[->, line width=1mm] (s1) edge[bend right=20] node[above, sloped] {deposit} (b3);
\draw[->, line width=1mm] (x2) edge[bend right=20, dashed] node[above, sloped] {
    report
} (b2);
\draw[<->, line width=1mm] (x2) edge[bend left=20, color=red] (r4);
\end{tikzpicture}}
\end{center}

\caption{\cz{Schematic representation of an operating model for a CBDC system.}
The diagram depicts the circulation of digital assets, interaction among
actors, and supporting functions.}

\label{f:operating-model}
\end{figure}

We present a prescriptive model for how to use our architecture to implement
CBDC, explicitly highlighting how CBDC would operate within the context of a
modern banking system and institutions.  We observe that money constitutes a
complex system within an economy, entailing a delicate set of connected
relationships among participants.  Our proposed architecture avoids undermining
this balance of connected relationships by aligning closely to the system
architecture implicit to physical cash.  In this sense, what we propose is not
a radical new system design, but rather a new kind of digital cash that can
exist alongside physical cash and other forms of money or money-like
instruments used for payments.  To support this model, we must consider the
processes and institutions that support the circulation of cash and how they
would be adapted to support the circulation of CBDC.  We also introduce two new
systems: an \textit{integrity system} comprising the set of relays, which
ensures that digital assets can be safely used to transfer value, and a
\textit{monitoring system} comprising the set of minters, for controlling the
creation and destruction of currency tokens.  Figure~\ref{f:operating-model}
illustrates how this would work, and we offer the following narrative
description of the lifecycle of a specific CBDC asset:

\begin{itemize}









\item \cz{Act I.} A unit of CBDC begins its life as a request from Alice to her
commercial bank, which had previously received a set of CBDC vouchers from the
central bank in exchange for reserves of equal value.  CBDC vouchers are
special CBDC assets that can be exchanged for signatures from minters but are
not used by retail consumers.  Alice's bank debits the value of the request
from Alice's bank account and sends the CBDC voucher to the minter along with
Alice's request.  The minter then signs Alice's request, destroys the voucher,
and submits a record of its work to the distributed ledger of the monitoring
system, which the central bank and regulators can inspect to understand the
aggregate flow of money in the system and verify that the minting invariant is
maintained.  The minter then sends the signed request back to Alice's bank,
which forwards the signed request to Alice.

Later, Alice uses the signature to create the CBDC asset, which we shall call
Bill, and transfers it to Bob.  Whenever a CBDC asset changes hands, either the
sender or the recipient must send an update to the correct relay to consummate
the transaction.  Next, Bob transfers Bill to his bank.  Importantly, unlike
Alice, Bob can execute this transfer immediately if he chooses to do so; there
is no particular value in waiting.  At the same time, unlike with the system
proposed by Chaum, Grothoff, and M\"oser~\cite{chaum2021}, Bob can wait as long
as he likes (subject to optional conditions) before depositing the asset with
the bank, since there is no requirement for the issuer or a minter to
participate in the transfers.  Finally, Bob's bank credits the value of the
transaction to Bob's bank account.

\item \cz{Act II.} Soon afterward, Charlie, another customer of Bob's bank,
makes a request to withdraw CBDC.  The bank sends Bill to the minter to be
recycled in exchange for signing Charlie's signature request.  The minter
destroys Bill, signs Charlie's signature request, and returns the signature to
Charlie via the bank.

Later, Charlie uses the signature to create a new CBDC asset, Bill II, and
transfers the asset to Dave.  Dave then transfers it to his bank, as Bob had
done.  Dave's bank decides to bring Bill II back to the central bank in
exchange for reserves, instead of recycling it, ending the lifecycle of the
unit of CBDC.


\end{itemize}

Note that Dave's bank could have done what Bob's bank did and save the CBDC to
service future requests without vouchers.  This recycling process is adiabatic,
does not rely upon the active participation of the central bank, and can be
repeated an arbitrary number of times in this manner before the ultimate
destruction of the unit of CBDC.  The minting invariant ensures that the
minting system never increases or decreases the total amount of currency in
circulation.  Instead, it issues a new unit of currency only in response to
collecting an old unit of equal value.  The central bank is only involved when
it engages with banks, specifically by issuing vouchers or accepting CBDC
assets in exchange for reserves, and by overseeing the minting operation,
passively accepting and analysing reports by minters.  The central bank also
relies upon the relay system to maintain CBDC integrity, and the DLT system
underpins its ability to verify what it must trust.

Note also that Alice's bank could have accepted cash or CBDC assets instead of
an equal amount of value from her bank account, although legal or regulatory
restrictions applicable to the acceptance of cash or CBDC assets might apply.

Finally, Alice could have transferred money directly to Bob's bank account
rather than to Bob.  Depending upon Bob's preferences, this might be a better
choice.  For example, it would reduce the total number of relay requests,
correspondingly reducing the operating cost to the relay system and
communication overhead for Bob.  It would also allow Bob to handle the case in
which Alice does not have exact change; Bob could forward Alice's signature
request in the amount of her overpayment to his bank along with his deposit,
and then return the blind signature for Alice's change directly to Alice.

\subsection{Managing CBDC distribution}

The central bank would handle the issuance, expiry, and destruction of its CBDC, as
well as managing its value though monetary policy.  Meanwhile, one or more
clearinghouses or banks would handle all of the real-time processing.  As part
of the issuance process the central bank may allow one or more
clearinghouses or banks to provide signatures on blinded templates, to be used by their customers in the final step of CBDC creation.  
The central bank would issue a specific quantity of some currency by
explicitly allowing a clearinghouse or bank to create and distribute
signatures for making that many units of CBDC.


We introduce the idea of a \textit{minting-plate}, which combines a \textit{minting-key} 
that can be used to sign blinded templates with a set of rules that govern its use.  
There is a deep tension between the desire to limit the number of units that 
can be created with a particular minting-key, and the need to prevent specific
units of currency from being connected to particular creation events 
(i.e. disconnected creation <-- fix this with the right name).
Because there is no way to connect a particular unit of currency with a particular creation event, 
there is also no way to tell whether a particular unit of currency was created by a legitimate user
of a minting-key, as opposed to a compromised or malicious use of that minting-key.

What can be done is to keep a record of how many units have been reportedly created and how many have been redeemed.
Creation is reported primarily by delegated issuers who holds a minting plates, 
and secondarily by retail banks which channel requests to those delegated issuers. 
Redemption happens when a bank brings CBDC units back to the central bank 
in exchange for central bank reserves. 

Together these values can reveal that a particular minting-key has been compromised,
which can help limit the damage caused by such a compromise.
A minting-key might be associated with a set of parameters to
limit, for example, the value of currency signed by that minting-key that is
in-flight at any particular moment (issuance minus redemptions), the total
value of CBDC cumulatively signed by that minting-key, and the time
at which signatures by that minting-key would no longer be considered valid.

The size of the anonymity set, as we shall discuss later in this section, is directly impacted by the limits that can be specified for the minting-plate. As more limits are placed on a particular minting-plate, the amount of currency it can produce is reduced, making it easier for powerful entities to track the behaviour of individual users.  It is important to tune those parameters so they provide good risk mitigation in the event of the compromise of a minting-key, while still maintaining a sufficiently large anonymity set.

\subsection{Managing CBDC system integrity}

In addition to managing the lifecycle of the individual CBDC assets, we imagine that the central bank would also take responsibility for establishing the integrity system for those USO assets. This integrity system must continue operating without equivocation, and it is possible to build it in a way so that it would not be impacted by increases in the number of assets, users, or transactions.

As an example, the central bank may declare that only licensed clearinghouses may operate relays that connect directly into its integrity system. Commercial and retail bank relays would connect into those clearinghouses, and relays operated by other money service providers would connect into those, along with third party corporate relays. Because the trust requirements for operating a relay are quite low, similar to those for a network carrier, this provides a rich ecosystem on which consumers can rely with no increase to the operational overhead of the integrity provider system.

Because the scaling concerns are mitigated, there is room to deploy heavyweight solutions for governing this integrity system. While it could be run from a single laptop, it is clearly better to design a system that is as resilient as possible. This means bringing all of the participants in the ecosystem together, such that not only the central bank, but also clearinghouses, commercial banks, retail banks, and so on are participating in a federated or decentralised system, so that only some proportion of them have to be operating correctly for the system to maintain the integrity of its operations.

It is worth explicitly noting that the computational cost of decentralised systems generally stems from two sources: one is the gatekeeping cost of keeping out bad actors, which is the primary reason for the hashing cost of proof of work based systems like Bitcoin and Ethereum; the other is the scaling cost of accommodating transactions, assets, and accounts. 

Our proposed architecture eliminates both of these costs. The first is
eliminated by only inviting trusted parties to add their efforts to the
integrity system. The second by separating the integrity system from
maintaining the state of the assets themselves, so that the scaling costs are
not borne by the integrity system.  Introducing good governance and
transparency into the integrity of a system does not necessitate a large
increase in energy usage. Our architecture demonstrates this.




\subsection{Managing regulatory compliance}
\label{ss:regulatory}

Ensuring that regulators can perform their duties is clearly an extremely important aspect of a well-functioning economic system, and must be an explicit goal of any realistic CBDC proposal. 
As we show in this work, regulatory compliance does not have to come at the cost of sacrificing consumer protections.  Indeed, not only are regulation and privacy compatible, but our architecture actually allows them both to be achieved more efficiently than current solutions that choose one over the other.

We have two main techniques for ensuring consumer protections. The first is the use of USO assets, which allow the CBDC to be acted upon by its owners unilaterally, regardless of the disposition of the financial apparatus. This means that while the recipient can choose to reject a transaction, no one else in the system, including regulatory bodies, can block it from happening or discriminate against that user.

The second is unlinking the sender from the recipient in the transaction channel. This means that even a powerful entity that knows who withdrew CBDC and knows who deposited CBDC will not be able to match senders to recipients.

How is efficient regulatory compliance possible with strong consumer protections like these? 
There are four places that regulation applies in our CBDC architecture, and they mirror four cases in which regulation applies to the use of cash.  We argue that we can not only satisfy but actually improve upon the established compliance procedures in each case:

\begin{enumerate}

\item\cz{When a retail user deposits cash into a bank account.}  Banks are often required, for cash deposits greater than a certain size, to request evidence from depositors that the cash to be deposited was obtained legally.  From this perspective, CBDC implemented as USO assets is better than cash, because it is possible to automate not just the integrity checks but also the regulatory checks.

\item\cz{When a retail merchant receives cash from a consumer.}  When merchants decide to deposit cash that they have received in the course of their business activities into bank accounts, they generally have an interest in knowing that the cash they have collected will be accepted.  CBDC implemented as USO assets allows such a merchant to apply the same integrity and regulatory checks that are run by their bank.  For example, a regulator might want to associate each recipient of CBDC with a bank account for the purpose of implementing compliance procedures.  To satisfy this requirement, we might stipulate that banks must require the recipient of CBDC to furnish a commitment in the form of its bank account details to any sender from which it might receive CBDC, and that the CBDC must include a signature of this commitment from the sender as a prerequisite for the bank to consider the CBDC to be valid.

\item\cz{When a retail merchant spends cash that it has received.}  Recipients of CBDC might want to spend it immediately without depositing it first.  Because USO assets track their own history, the next recipient is able to know whether the CBDC has travelled around since leaving a bank.  Therefore, the asset must carry the burden of proving that its travel satisfies the relevant regulatory requirements, which could be enforced by automated checks run by the bank that ultimately receives it in the form of a deposit.  

In this manner, a regulator might allow CBDC to travel over multiple hops, with multiple recipients of CBDC in succession, without the interactive involvement of a regulated financial institution, provided that the recipient bank account details are included and signed by the respective sender in each successive hop.  Note that, although the first sender might be anonymous, the USO asset framework enables it to implicitly demonstrate its possession of the key signed by the issuer of the CBDC.  Subsequent senders would be identified by their bank account information as recipient from the previous transaction.  Conversely, a regulator might want to enforce a rule that recipients of CBDC can do nothing other than deposit CBDC that they receive directly into the specified bank account. To satisfy this requirement, we would stipulate that banks would enforce a rule that the USO asset must have been transacted no more than once (i.e., only one hop).

The rules are implicitly dynamic. Bob's bank chooses what program to run to conduct the automated regulatory check, and Bob's software uses the same program as Bob's bank, so regulators can change their requirements at any time without needing the issuance of new CBDC.  Regulators could do this by asking the banks to update their compliance procedures, and those new requirements would then be applied within the software of consumers and merchants.

\item\cz{Compliance procedures within a financial institution.}  A financial institution can prove that in all cases the CBDC it has accepted has met the current regulatory standards.  Either the asset passes the automated regulatory checks, or the institution has accepted external evidence to meet the regulatory requirement.  We imagine that the latter case would be extremely rare, because consumer and merchant software would automatically reject CBDC that does not meet the regulatory checks that would be carried out by their bank, but it provides an important safety valve.
    
\end{enumerate}

To achieve the desired regulatory protection, the source and sink of CBDC must
be regulated entities. When Alice creates new CBDC, the signature granting it
validity must come from a regulated financial entity; this is enforced by the
central bank or its delegates such as minters. When Bob brings his CBDC back to
a regulated financial entity such as his bank then that entity can return the
CBDC to the central bank in exchange for reserves.

Our architecture is compatible with a variety of additional mechanisms for
enforcing regulatory requirements, although we recommend careful consideration
to verify that such mechanisms are compatible with consumer protection
objectives such as privacy and ownership.  Note that the first transaction in
which a new asset changes hands provides consumer protection, although
subsequent transactions do not.  In particular, although the initial consumer
is protected, the merchant might decide to spend his or her CBDC asset in a
second transaction rather than have a bank recycle it, but he or she does this
knowing that what the second recipient does with the CBDC asset might expose
sensitive information about the second transaction.

Having regulated entities as the source and sink of CBDC is sufficient for a
mechanism to ensure full regulatory compliance. More than this, it allows that
compliance to be achieved with widespread efficiency gains: for the regulator,
for the banks, for merchants, and for consumers.

\subsection{Ensuring an appropriate anonymity set}

In our formulation, CBDC is generally not held by retail customers in custodial
accounts and, for this reason, would not earn interest.  Although there are
some methods available by which fiscal policy can incentivise or disincentivise
spending tokens~\cite{goodell2020}, we expect that retail users would view CBDC
primarily as a means of payment rather than a store of value.  We stipulate
that plausible deniability is essential to privacy, and a large anonymity set
is a prerequisite to plausible deniability.  Inexorably, a trade-off between
privacy and flexibility for users lies in the relative timing of withdrawals and
remittances, as the strength of the anonymity set is bounded by the number of tokens in-flight between those events.

The template architecture ensures that the consumer chooses the minting-key.
We assume that the set of minting-keys signed by the issuer will be available
for public perusal on a distributed ledger.  The fact that an issuer cannot
sign multiple minting-keys without having that fact become observable forces
accountability for an issuer that might want to create a covert channel that
could reveal information about the consumer.  Since retail users would have no
particular reason to hold CBDC longer than is necessary to make their payments,
just as they would have no particular reason to hold cash, it is important to
consider ways to encourage users to hold CBDC long enough to ensure that the
anonymity set is large enough to protect their privacy.  In service of this objective, we propose some practical mechanisms that can be applied to ensure that the anonymity set is sufficiently large to protect the privacy of everyday users:

\begin{itemize}

\item\cz{Encourage consumers to withdraw larger amounts of money.}  For
example, consumers can withdraw CBDC in fixed-size lots, and then spread out
the use of those over a longer time period and blend in with other consumers,
thereby making a smaller number of larger-sized withdrawals from the bank.  We
anticipate that reducing the number of withdrawals will make it harder to link
a payment to its corresponding withdrawal, potentially by one or more orders of
magnitude.  By reducing the number of statistically linkable withdrawal-payment
pairs, users can enjoy a larger anonymity set and, as a result, better privacy.

\item\cz{Incentivise consumers to use slow relays by default.}  We can give
users control over the extent to which it might be possible to temporally
correlate a withdrawal to the proof data that is created with a payment.  This
can be accomplished by adjusting the requirements in Step 4 of the user
engagement lifecycle (refer to Figure~\ref{f:aliceflow}) such that $F_1$ can
only be accepted by relay $G$ if $F_0$ had previously been published by relay
$G$.  Then, relay $G$ can explicitly specify a frequency for its publication of
successive updates to ensure a sufficiently large anonymity set, for example, to
publish once per minute, hour, or day.

The motivation is to increase the cooling off period to increase the number of
unspent withdrawals from the same minting-key.  The provider of relay $G$ could
maintain multiple relays with different frequencies.  If we accept privacy as a
public good~\cite{fairfield2015} and acknowledge transaction immediacy as a
threat to privacy, then the provider could charge more to consumers who demand
greater immediacy, as a way of compensating for the negative externalities that
would result from shorter time intervals between withdrawals and payments.
Since the consumer's message to the relay requires no human interaction, CBDC
software could send it after a random delay, or could send it through a
remailer network such as Mixmaster~\cite{mixmaster}.

\item\cz{Encourage slow transaction settlement when possible.}  Not every
transaction must be settled immediately; consider the case of online purchases
for goods or services to be delivered in the future.  For such transactions, if
Alice can use Step 4, Option 1 (as shown in Figure~\ref{f:step-4-1}) to give Bob
direct control and the means to acquire possession of the CBDC, and if Alice
trusts Bob not to record the time at which she does so, and if Alice trusts Bob
to delay his request for the proof of provenance (and thus settlement) for a
sufficiently long time, then Alice can effectively pay Bob immediately.
Indeed, Bob's transaction tracking and rate of transactions might influence
Alice's calculations about whether this option is safe.  Note that this is the
same guarantee that payers rely upon to safely use physical cash without being
tracked.  In the digital context, procuring a strong guarantee about what Bob
might do is somewhat harder, and we are pessimistic about the idea that
received transactions are not being timestamped, either by Bob or by other
observers.

\item\cz{Have Alice explicitly give control to Bob during the withdrawal phase.}  Alice can give control to Bob in the creation of $F_0$ during Step 1 of the protocol.  Because $F_0$ is part of the blinded template, neither her bank nor other observers will be able to associate her withdrawal with her payment to Bob.  As with the previous approach, this approach requires Alice to trust Bob not to record the time at which he receives the payment from Alice.  However, because Bob is able to verify that the CBDC is valid and that he has exclusive control, this approach might be appropriate for immediate delivery of goods or services.  Although the size of the transaction might ordinarily reveal information that could link the withdrawal to the payment, this could be obfuscated by having Alice give Bob a larger quantity of CBDC than he requires, and having Bob provide Alice the excess in the form of new CBDC, either immediately or in the future, using the same method.




\end{itemize}

We also suggest implementing a mechanism to monitor the number of tokens
currently in-flight, to support dynamically adjusting parameters that could
impact the size of the anonymity set, such as the number of minting-keys, the
number of tokens to be issued by each minting-key, and the set of available
denominations.  Such a mechanism would support not only the management of
digital currency issuance and destruction but also public oversight of the
entire process.

\subsection{Clearing and settlement}

Ensuring that the integrity system continues to produce entries and does not
equivocate about the history of is commitments is a major responsibility of a
central bank that produces CBDC using this architecture.  This can be done by
the central bank directly, although such an approach introduces a set of risks,
including the possibility that the central bank's operational servers crash or
become compromised as well as the possibility that the central bank might
change the rules or expectations for the system without warning.  Because
distributed ledgers are designed to be fault-tolerant and immutable, DLT is a
useful tool for systems that require some resilience to crashes and compromise.
We suggest that the central bank could take the following approach to using DLT
for its integrity system:

\begin{enumerate}

\item The central bank enlists several highly trusted but independent
institutions to run relays and requires each of them to sign off on each new
entry that the central bank produces.  This protects against compromise of the
central bank: The adversary must also compromise all of the other institutions
to cause an equivocation.

\item The institutions employ a crash fault tolerance mechanism, such as
Raft~\cite{ongaro2014}, to allow a few institutions to be offline without
interrupting the operation of the system.

\item The institutions themselves can propose new entries, perhaps via a fixed
schedule or round-robin process, instead of requiring the central bank to do
it.  This avoids issues associated with having the central bank serve as
gatekeeper to transactions and allows the central bank to step out of an
operational role and focus on oversight and governance.

\item The institutions make a commitment to publish every entry they sign.

\end{enumerate}

This arrangement is sufficient to convert the centralised integrity
system
into a distributed ledger overseen, but not operated, by the central bank.

The scalability of this architecture can be enhanced by allowing relays to
arrange themselves hierarchically.  Higher-level relays can aggregate the
entries produced by lower-level relays and perform the same process, with the
respective lower-level relay operators taking the place of the trusted
institutions.  Waiting for a higher-level relay to produce an entry might
support greater assurance that the proof will be completed, but might be slower
than waiting for the lower level relays, which are optimised to minimise
latency.

Transactions less than a specified amount might be considered final by
transacting parties, and may be covered by appropriate insurance or credit for
relay operators, without confirmation from the clearing network.  The
additional confidence provided by aggregate confirmations, therefore, might be
necessary for buying high-value goods, such as a car, but probably not for
buying low-value goods, such as a cup of coffee.

A case can also be made for encouraging relay operators to use mechanically
external DLT systems as a commitment mechanism, or public bulletin board, for
publishing their entries.  This practice might also enhance the confidence in
those entries, as well as quicker detection of equivocation of compromised
relays, because it compels relays to commit to a more unified view of their
published entries rather than merely self-reporting them.

\section{Use cases}
\label{s:use-cases}

In this section, we consider three use cases that demonstrate the power and
flexibility of our design and how our proposed architecture can be used to
satisfy them.  These use cases offer advantages over other electronic payment
methods, including modern retail payments via banks or payment platforms as
well as unlinkable CBDC proposals such as the one offered by Chaum, Grothoff,
and M\"oser~\cite{chaum2021}.  The users of the system, including consumers and
service providers, can choose which of these possibilities to enable and
support.

\subsection{Disconnected operation}

In some environments, access to the central bank might be slow, delayed, or
intermittent rather than real-time, for example where the central bank might be
accessible only at certain times.  
We refer to such environments as ``disconnected'',
and we imagine that this characteristic might apply to some remote or
sparsely-populated areas with limited or unreliable connectivity, as well as
categorically isolated environments such as certain remote villages, ships in 
the high seas, aircraft in flight, spacecraft in space, or remote military outposts.

Fair exchange requires the involvement of a mutually trusted third party~\cite{pagnia1999}.  However, this does not imply that
all transactions must take place with global agreement.  In disconnected environments
we assume that there exists a local actor who is sufficiently trustworthy to act as a relay
for nodes within that environment. This might be a trusted institution, a network operator,
or even a distributed system made up of the nodes in that environment.

As long as the recipient trusts that relay to not equivocate, then the recipient can accept a payment that
has a proof of provenance that includes that relay, with confidence that it will be possible to complete 
the proof of provenance to include the integrity provider. Completing that proof is necessary for the 
payment to be accepted outside of the environment in which the relay is trusted to do its job, 
but inside of this environment payments can continue to be made without making external network connections.
As long as the trusted relay does not equivocate, then nothing that anyone else does, either inside the environment
or outside, can adversely impact the payment.  Short of equivocating, nothing the trusted relay does, 
including crashing or denying service, can adversely impact it either.

We note that systems that require global consensus, including all centralised
systems and most distributed ledger systems, lack this capability.



\subsection{Offline operation via time-shifting}
\label{ss:time-shifting}

Some environments have no connectivity at all.  This might include environments
without communication equipment, or environments without a local point of
trust.  We refer to such environments as truly ``offline''.  Since transactions
require a third party~\cite{pagnia1999}, it might seem that this means that
offline transactions are impossible, but that is not entirely true.  The
involvement of the third party could take place at a different point in time.

A user can transfer CBDC to an address over which the recipient has control, but without revealing to the recipient the information needed to exercise that
control.  
Then the user can then effectively spend the CBDC offline by revealing information about the transfers to the recipient.  
In the event that the user decides not to spend
all of the CBDC with that recipient, they have the option to use a fair-exchange protocol with the recipient to redeem any CBDC that was
transferred but not spent.

In principle, it would be possible to transfer CBDC to a market operator in exchange for tickets (perhaps implemented using blind signatures)
and then give the tickets to merchants, and the
merchants could use a fair-exchange protocol to redeem value from the market
operator.  However, this assumes that the merchants are connected to the
market operator in real-time so they can verify that
such tickets are still available to claim.  Similarly, it might be possible to transfer CBDC to an issuer of cash-like, counterfeit-resistant physical tickets that can be used in a local context to make offline purchases to arbitrary recipients without the need for a real-time network connection.

\subsection{Chained transactions with embedded provenance}
\label{ss:chained}


There are several reasons why a recipient of CBDC might want to move it onward
without depositing the CBDC directly into a bank account.  We refer to such
transactions as \textit{chained transactions}.  In such cases the provenance information about successive holders of an asset can be
maintained within the CBDC tokens, and chained transactions can carry their own
proofs of compliance with the rules of the system.  Appropriate use cases might
include the following:

\begin{itemize}

\item Perhaps a CBDC holder has no access to a bank or access to a bank
is difficult as a result of network connectivity or geographic location.  Being able to make a series of transactions under such circumstances may provide an important safety net. 

\item Perhaps a CBDC holder is acting on behalf of a business that seeks to
maintain provable records of its internal or external transfers, perhaps to
streamline compliance operations, to satisfy auditing requirements, or to move
assets without depositing them into a bank account and incurring a delay
associated with settlement.
For example, a multinational corporation might want to preserve
an audit trail of internal transactions, for example to demonstrate compliance
with tax regulations concerning the applicable jurisdiction for revenue, in
addition to economic efficiency for such internal moves.

\end{itemize}

\section{Analysis}
\label{s:analysis}

In this section, we compare our architecture to alternative architectures for exchanging value.  We begin with a set of mechanical design choices and argue for the choices inherent to the argument that we have proposed.  Then, we compare our architecture to other systems for exchanging value in terms of the asset-level requirements and system-level requirements defined in Sections~\ref{ss:asset} and~\ref{ss:system}.

\subsection{Comparison to other untraceable CBDC solutions}


Chaum, Grothoff, and M\"oser~\cite{chaum2021} have also proposed a system for
untraceable CBDC.  Our system also leverages the blind signature mechanism that
is central to their design, although our system differs from theirs in several
important ways.  In particular, our system:



\begin{itemize}

\item \cz{Enforces accountability and transparency for authorities and system operators} by leveraging distributed ledger technology as described by Goodell, Nakib, and Tasca~\cite{goodell2021}, thus requiring authorities or system operators to explicitly and publicly specify changes to the protocol and system rules;

\item \cz{Enables transactions without real-time involvement of the central
bank or issuing authority}, by progressively, and obliviously, building proof structures with logarithmic scaling factors across the relays; and

\item \cz{Enables validations without any involvement of the central
bank or issuing authority}, by incorporating self-validating proofs of provenance as a fundamental part of the digital assets; and

\item \cz{Avoids requiring the central bank to maintain a database} of
individual tokens, balances, or specific transactions, as is done with
UTXO-oriented digital currency systems.

\end{itemize}

\subsection{Design features}

Some of the design features of our proposal distinguish it from alternative proposals available in the current literature on digital currency.  We list several of the most important such features here:

\begin{itemize}

\item\cz{Regulatory control applies to transactions, not asset ownership.}
Our proposed architecture allows regulatory compliance to be automatically enforced by regulated financial institutions that receive CBDC on behalf of their account-holders.
This allows comprehensive regulation without introducing a requirement to track the ownership of every token.

\item\cz{Non-custodial wallets.} People want custodial accounts because they want strong regulatory controls. Having strong regulatory control at the transaction level allows non-custodial wallets to operate within the regulatory regime, providing efficiencies that make more use cases available to the users of CBDC.  This approach allows CBDC to realise the benefit of a token-based approach, while interoperating with traditional custodial accounts as desired, as cash does. 

\item\cz{Open architecture.} Our approach does not rely upon trusted computing, including trusted software, trusted hardware, or secure elements of any kind.  Device manufacturers are third parties, just as other authorities are, and requiring any trusted authority to be part of every transaction compromises the integrity of the system.  This is important because we do not wish to require the establishment of a set of trusted hardware vendors, or the assumption that counterparties to a transaction must trust each other's devices.  If counterparties do have mutual trust in a third-party, such as an institution, they can use this mutual trust to improve the efficiency of a transaction, as described in Section~\ref{ss:effset}.

\item\cz{Time-shifted transactions.} Because fair exchange always requires a third party to every transaction~\cite{pagnia1999}, we observe that there is no way for two counterparties to transact directly without access to a mutually-trusted third party or system.  
In cases where a mutually trusted system is inaccessible, our architecture allows a 
time-shifted trust in the form of prepayments, as described in Section~\ref{ss:time-shifting}.

\item\cz{Decentralised transactions.}  By allowing transactions to be processed in a decentralised manner, our approach avoids the costs and risks of requiring a ledger or other system component to be under the control of a single actor, who might change the rules without public oversight, discriminate against certain users, equivocate about the history of transactions, or otherwise exercise arbitrary authority.

\item\cz{Energy efficiency.}  By allowing transactions to be processed locally, our approach avoids the costs and risks of requiring a heavyweight, ledger-based system (distributed or not) to be in the middle of every transaction, allowing the use of the CBDC to be highly energy efficient.

\item\cz{No central user database.}  Our system avoids introducing centralised identity requirements, leveraging the existing decentralised procedures for identification and compliance that are already widespread among financial market participants.  This avoids establishing new mechanisms to track users and aligns with global agreements about compliance requirements.

\end{itemize}

\subsection{Efficient settlement}
\label{ss:effset}

One of the most important features of cash infrastructure is the ability of
counterparties to transact in real-time, with minimal involvement of third
parties.  To the extent that third parties are not involved in transactions,
they cannot engage in rent-seeking behaviour and cannot pass the costs they
incur along to transaction counterparties in the form of fees.  Where third
parties are involved, the involvement is generally minimal and highly local,
for example to provide cash withdrawal services (e.g. ATM infrastructure) for
consumers and cash deposit services to merchants, both of which are used only
in aggregate over many transactions.  Cash infrastructure also benefits from
instant settlement: Once a payer has given cash to a payee, the transaction is
settled.  There is no way for a payer to unilaterally unwind (``claw back'')
the transaction.

With modern digital transactions, scalability interferes with the ability to
transact in real-time.  Transactions take place across a network, which cannot
be globally synchronised.  Settlement requires pairwise synchronisation between
transacting institutions, which must manage risks associated with concurrency.
Settlement times for domestic bank wires and direct debits are generally a
matter of hours; settlement times for international wires are even longer.
Payment networks generally offer short-term credit as a way to support faster
settlements.

Our system design provides a mechanism for two transacting parties to enjoy
real-time settlements.  Recall that, in general, a payer (Alice) must furnish a
proof of provenance to a payee (Bob) before a payee will accept payment,
and that Alice creates this proof by connecting to the issuer through her chosen relay.
If Alice is always assumed to be directly connected to the issuer, then the system will not scale very well: the issuer would have a de facto role in every transaction, and the resulting need to serialise and batch transactions would mean that Alice might be forced to wait.

However, because a payer can choose the relays, Alice has the option to choose one that both she and Bob recognise as trustworthy.  
Because each of these relays is a checkpoint in building the proof of provenance 
they can offer guarantees to Bob that Alice's transaction has been incorporated.  If Bob trusts a relay that Alice has chosen, then this partial proof of provenance will suffice until Bob has received the full proof of provenance.

Our architecture allows these promises to be made almost instantaneously by these relays, requiring very little computation. Additionally, various mechanisms can be used to reduce the risk that a relay would equivocate by rewriting history to nullify Alice's transaction.  
These include both traditional institutional and legal guarantees as well as technical mechanisms like distributed ledgers and other means of achieving immutability.

If Alice knows that she is likely to make a
purchase within a context in which a particular relay is trusted, then
Alice can choose to use that relay for her asset, thus allowing near
real-time payments within that context.

We observe that this mechanism offers similar functionality to debit card transaction via a retail payment network, wherein transactions can be accepted in real-time because the retail payment network provides a guarantee to the recipient's financial institution that the transaction will succeed.  Our proposed mechanism avoids some of the potential friction intrinsic to this approach by eliminating the need for financial credit, although Bob must trust the relay to fulfill its promise to incorporate the transaction.  Additionally, because transactions involve direct obligations of the central bank rather than bank deposits, the requirement for a clearinghouse to resolve counterparty risk among institutions is eliminated.



\subsection{The fallacy of anonymous accounts}

There are two chief approaches to mitigating harmful consumer tracing and
profiling.  One approach is anonymous accounts, where the identity of the
account holder is decoupled from the account.  Anonymous accounts are akin to
prepaid debit cards and have been proposed as a way to protect the rights of
consumers~\cite{pboc2021}.  The other approach is transactional unlinking,
wherein the sender is decoupled from the receiver inside the transaction
channel.  These two approaches of anonymous accounts and transactional unlinking
are actually orthogonal dimensions.

In the absence of transactional unlinking, anonymous accounts don't provide
anything useful.  Bitcoin is a stark example of this: regular transactions can be
trivially de-anonymised, revealing a consumer's entire history, whereas
criminals can employ various heavyweight measures to conceal themselves.

In the presence of transactional unlinking, anonymous accounts still don't
provide anything useful: the transactional unlinking already stops unwanted tracing and
profiling, and adding anonymous accounts on top of that only makes enforcing
regulatory compliance much more difficult.

Thus, we conclude that anonymous accounts are worse than useless.  They do not
achieve their stated goals, and they extract a high cost from systems that
employ them~\cite{goodell2021a}.  We also note that anonymous accounts
typically contravene AML/KYC recommendations and, because they implicitly link
successive transactions done by a consumer to each other, are not actually
private for most legitimate retail use.

We assume that the accounts referenced by our system would be subject to
AML/KYC data collection and would not be anonymous.  The privacy of our
approach results from the use of non-custodial wallets to unlink successive
transactions involving the same currency.  Specifically, a user must
``withdraw'' funds from a regulated money services business into her
non-custodial wallet in one transaction and then ``remit'' funds into a
regulated money services business in the next.  Even though the holders of the
payer account and the payee account are known, the fact that money has flowed
between them is not.

\subsection{A comparison of payment system architectures}

\begin{table}[ht]
\begin{center}

\sf
\begin{tabular}{|L{9.3cm}|p{\ccol}p{\ccol}p{\ccol}p{\ccol}p{\ccol}p{\ccol}|}\hline
& \rotatebox{90}{Cash}
& \rotatebox{90}{Custodial accounts}
& \rotatebox{90}{Traceable digital currency}
& \rotatebox{90}{Untraceable digital currency}
& \rotatebox{90}{Traceable USO digital currency\,\,\,\,}
& \rotatebox{90}{Untraceable USO digital currency\,\,\,\,}\\
\hline\textbf{Integrity Considerations}                     & & & & & & \\
Durability                                                  & \CIRCLE & \Circle & \Circle & \Circle & \CIRCLE & \CIRCLE \\
Self-contained assets                                       & \CIRCLE & \Circle & \Circle & \Circle & \CIRCLE & \CIRCLE \\
\hline\textbf{Control Considerations}                       & & & & & & \\
Mechanical control                                          & \CIRCLE & \Circle & \Circle & \CIRCLE & \CIRCLE & \CIRCLE \\
Delegation                                                  & \Circle & \Circle & \Circle & \Circle & \CIRCLE & \CIRCLE \\
\hline\textbf{Possession Considerations}                    & & & & & & \\
Choice of custodian                                         & \CIRCLE & \Circle & \Circle & \Circle & \CIRCLE & \CIRCLE \\
Choice to have no custodian                                 & \CIRCLE & \Circle & \CIRCLE & \CIRCLE & \CIRCLE & \CIRCLE \\
\hline\textbf{Independence Considerations}                  & & & & & & \\
Fungibility                                                 & \CIRCLE & \Circle & \Circle & \CIRCLE & \Circle & \CIRCLE \\
Efficient lifecycle                                         & \Circle & \Circle & \Circle & \Circle & \CIRCLE & \CIRCLE \\
\hline\end{tabular}
\rm

\caption{A comparison of payment system architectures by asset-level considerations.}

\label{t:comparison:asset}
\end{center}
\end{table}

\begin{table}[ht]
\begin{center}

\sf
\begin{tabular}{|L{9.3cm}|p{\ccol}p{\ccol}p{\ccol}p{\ccol}p{\ccol}p{\ccol}|}\hline
& \rotatebox{90}{Cash}
& \rotatebox{90}{Custodial accounts}
& \rotatebox{90}{Traceable digital currency}
& \rotatebox{90}{Untraceable digital currency}
& \rotatebox{90}{Traceable USO digital currency\,\,\,\,}
& \rotatebox{90}{Untraceable USO digital currency\,\,\,\,}\\
\hline\textbf{Autonomy Considerations}                      & & & & & & \\
Privacy by design                                           & \CIRCLE & \Circle & \Circle & \CIRCLE & \Circle & \CIRCLE \\
Self-determination for asset owners                         & \CIRCLE & \Circle & \Circle & \CIRCLE & \Circle & \CIRCLE \\
\hline\textbf{Utility Considerations}                       & & & & & & \\
Local transactions                                          & \CIRCLE & \Circle & \Circle & \Circle & \CIRCLE & \CIRCLE \\
Time-shifted offline transactions                           & \CIRCLE & \Circle & \Circle & \Circle & \CIRCLE & \CIRCLE \\
Accessibility                                               & \CIRCLE & \Circle & \CIRCLE & \CIRCLE & \CIRCLE & \CIRCLE \\
\hline\textbf{Policy Considerations}                        & & & & & & \\
Monetary sovereignty                                        & \CIRCLE & \Circle & \CIRCLE & \CIRCLE & \CIRCLE & \CIRCLE \\
Regulatory compliance                                       & \Circle & \CIRCLE & \CIRCLE & \CIRCLE & \CIRCLE & \CIRCLE \\
\hline\end{tabular}
\rm

\caption{A comparison of payment system architectures by system-level considerations.}

\label{t:comparison:system}
\end{center}
\end{table}

\noindent Tables~\ref{t:comparison:asset} and~\ref{t:comparison:system}
summarise the characteristics of a selection of different payment system architectures,
including our proposed architecture.  The descriptions of the payment
mechanisms are as follows:

\begin{itemize}

\item\cz{Cash.} A central bank produces physical bank notes and coins.  Retail
users circulate them freely, without involving of financial intermediaries.
Cash is part of the monetary base of an economy; commercial banks can exchange
cash for deposits with the central bank.  Although bank notes have serial
numbers, cash remains fungible because it can be freely exchanged among bearers
and because retail users of cash generally do not maintain records that
identify individual units of cash.

\item\cz{Custodial accounts.}  These are retail payments that take the form of
transfers between financial institutions.  This category covers both the case
of private-sector banks offering accounts to retail consumers as well as the
case of central banks offering accounts to retail consumers.  Such payments
might include bank wires, ACH, cheques, direct debit, and third-party transfers
via payment networks including but not limited to card payment systems. 

\item\cz{Traceable digital currency.} Retail consumers hold tokens that are obligations of the
central bank.  The tokens are bearer instruments and are not held in custodial
accounts, although individual tokens can be linked to the identities of their
owners.  Thus, the consumers are not anonymous and are therefore subject to
profiling and discrimination on the basis of their transactions.  The issuer
must maintain a record of tokens that were spent to prevent double-spending.
The record of tokens can be maintained by the issuer directly or by a
distributed ledger using a decentralised consensus system.

\item\cz{Untraceable digital currency.} This approach is similar to traceable digital currency, 
except that the central bank signs blinded tokens using a blind signature scheme of the
sort elaborated by David Chaum~\cite{chaum2021}.  When a user wants to spend a
token, the user unblinds the token and returns it to the issuer along with the
address of the recipient.  Recipients could be anonymous, or not anonymous,
depending upon the specifics of the architecture.  Chaum's proposal for digital currency
implicitly assumes that the sender is anonymous, but the recipient is not
anonymous in the usual case~\cite{chaum2021}.

\item\cz{Traceable USO digital currency.} This approach to digital currency uses
baseline USO assets.  The tokens are not blinded, and although tokens
can be directly transferred between possessors without the involvement of the
issuer, the chain of custody of an asset is transparent and completely
traceable to its possessors.

\item\cz{Untraceable USO digital currency.} This approach to digital currency is
a fusion of USO assets and the Chaumian system.  A user approaches
an issuer with a request for a blinded token, which the issuer furnishes to the
user.  When the user wants to spend a token, the user unblinds the token,
incorporates it into a specific previously created asset, and transfers
the asset to the recipient.  It is now up to
the recipient to redeem the token with the issuer, or to pass it to another
recipient without the benefit of anonymity.

\end{itemize}

\section{Conclusion}
\label{s:conclusion}

In this article, we have presented an untraceable version of an architecture for a payment system based on proofs of provenance.  Our architecture combines two previous lines of work to provide a solution that efficiently provides both consumer protections and regulatory compliance. Doing this allows the resulting CBDC to be used across a wide variety of use cases, including many of those currently addressed by cash.

Our proposal directly addresses the dilemma of maintaining regulatory compliance while preventing abusive profiling that harms consumers.  Abuses of profiling are endemic to modern payment systems, wherein not only governments but also consumer-facing businesses, service providers, and platform operators actively analyse consumer behaviour and can exploit personal information for profit or control.  Ordinary consumers are forced to trust not only the practices and motives of such actors but also their security. The costs and risks of security breaches are generally borne by the consumer, and can be quite severe: in the US alone they are estimated to amount to US\$228B in the last year~\cite{maler2021}.  Central banks have an opportunity to repair trust between citizens and the state by sponsoring an architecture that does not force users to trust some third party with data protection, but instead allows users to verify for themselves that their privacy is protected.

Solutions that promise to end profiling generally do so either by allowing anonymous accounts or by facilitating the consummation of transactions outside of an environment wherein regulators can operate and effectively supervise activity.  In contrast to such approaches, our proposal allows effective regulatory supervision while unlinking users' banking relationships from their spending habits, thus enabling consumers to enjoy fully regulated custodial accounts while avoiding the costs and risks of abusive profiling.

Furthermore, this architecture addresses concerns about the transactional efficacy of other untraceable architectures by allowing the recipient to accept payment without involving the issuing bank or the sender's financial institution.  It also provides a framework for strong assurance of provenance and auditability, allowing follow-on transactions to occur prior to involving the recipient's financial institution.

Our proposal addresses the operational and infrastructural overhead that a central bank must incur to manage a payment system through a domestic retail digital currency.  It provides an efficient path to the issuance and distribution of a currency as well as the maintenance of its integrity.  The distribution and management following issuance can be mediated by existing robust payment channels, including clearinghouses, commercial banks, and payment services businesses, using existing payment mechanisms and avoiding the costs and risks associated with deploying new infrastructure for that purpose.

This architecture also addresses the governance and risk mitigation concerns of issuing a domestic retail digital currency and managing a payment system by isolating the components of the system so that each can be treated independently, including the desired properties related to integrity, possession, control, and autonomy as well as the operations of issuance, distribution, and transaction management.  Our proposal thus encourages working within the current banking system, including commercial banks and payment institutions, rather than undermining them, and provides the capacity to build a deep and resilient governance approach without compromising the efficiency and privacy of individual transactions.

Cash is used in many different situations, as are other payment service solutions. We describe the properties a CBDC must have in order to be efficiently used in those situations, and we show that the technical requirements of our architecture are necessary to deliver a solution with those properties. This allows the CBDC created using our architecture to broadly meet the demands of cash as well as those of electronic payment services, and highlights exactly where other proposals fall short. It is not necessary to make unacceptable compromises between consumer protections and regulatory compliance, and it is not necessary to sacrifice operational efficiency to maintain asset integrity. Indeed, for a currency to be used like cash, it must excel in all three of those aspects. Ours does.






\section*{Acknowledgements}

The authors are grateful to TODAQ for sponsoring this work.  We thank Professor Tomaso Aste for his continued support for our project.  We also acknowledge the support of the Centre for Blockchain Technologies at University College London and the Systemic Risk Centre at the London School of Economics, and we acknowledge EPSRC and the PETRAS Research Centre EP/S035362/1 for the FIRE Project.

\end{document}